**Diameter dependence of light absorption in GaAs nanowires evidenced by photoluminescence spectroscopy**

*Francisca Marín*, Ijaas Mohamed, Oliver Brandt, and Lutz Geelhaar*

F. Marín, I. Mohamed, O. Brandt, L. Geelhaar
Paul-Drude-Institut für Festkörperelektronik, Leibniz-Institut im Forschungsverbund Berlin e.V., Hausvogteiplatz 5–7, 10117 Berlin, Germany
E-mail: marinlargo@pdi-berlin.de

I. Mohamed
Department of Electrical and Computer Engineering, The Ohio State University, Columbus, Ohio 43210, USA

Funding: We acknowledge funding from the Deutsche Forschungsgemeinschaft (DFG) under grant GE 2224/5-1.

**Abstract: Semiconductor nanowires are attractive for photovoltaic applications because light absorption can be enhanced compared to planar layers due to the more complex coupling of light with wavelength-scale matter. However, experimentally it is very challenging to measure light absorption in single nanowires. Here, we employ photoluminescence spectroscopy as a new method to investigate how the diameter of highly phase-pure GaAs nanowires affects light absorption. The underlying concept is that the absorption of the exciting laser light influences the photogenerated carrier density and in turn spectral features. In particular, we exploit that both the saturation of a specific defect line and the transition from excitonic to electron-hole-plasma recombination occur at well-defined carrier densities. We find that absorption is maximized for a diameter of about 80 nm. Our approach may be transferred to other material systems and thus enables straightforward systematic experimental studies of absorption enhancement in single nanowires.**

## 1. Introduction

In semiconductor nanowires (NWs), light absorption can be enhanced compared to what would be expected just for the given volume of material. One reason is that the subwavelength size of NWs gives rise to Mie resonances to which the incident light can couple more effectively and thereby enhance the light-matter interaction. These resonances are associated with the eigenmodes or leaky mode resonances of the NWs, which govern the optical properties, including scattering and absorption. These leaky mode resonances are determined by the NW diameter, the refractive index, and the wave vector of the incident light[1]. By engineering these parameters, it is possible to optimize the light absorption properties of the NWs.

In addition, careful design of NW arrays, i. e. NW diameter, length, and spacing, can suppress reflection and transmission while enhancing the scattering of incident light inside the array. In other words, light is trapped in the NW array which increases the optical path length, leading to enhanced light absorption [2-4]. These possibilities to engineer light absorption through NW and NW array design make these nanostructures an attractive platform for photovoltaic and sensing applications [5,6].

The investigation of light absorption enhancement in NWs and NW arrays often involves theoretical methods. Numerical finite difference time domain (FDTD) simulations of resonant absorption in periodic NW arrays have been carried out for varying illumination angles and array density[7], NW diameter[8], as well as NW shape and diameter randomness[9,10]. Other approaches include transfer matrix methods to optimize NW geometry[11] and NW array design[12], as well as analytical models as an alternative to time-consuming numerical simulations[4]. However, in any kind of theoretical approach, the details of the actual light absorption configurations such as realistic NW geometry, laser spot size, and scattering angle are often not taken into account [13-16], and therefore experimental measurements are a must.

Experimentally, light absorption in semiconductor NW arrays is typically determined via measurements of reflection and transmission using an integrating sphere, enabling thus to study the effect of array design [2,17]. Experiments with individual NWs are much more demanding since both a high spatial resolution and sensitivity are needed to investigate light absorption on the nanoscale. Most reports are based either on extinction spectroscopy [18-21] or photocurrent measurements [13-15,22]. In other cases, integrating sphere microscopy [23] and measurements of the heat flow under laser illumination [16] have been employed. Some of these approaches have achieved good sensitivity and others good spatial resolution. Nevertheless, all of them exhibit one or multiple significant experimental challenges, such as discerning contributions specific to the NW from environmental factors, intricate procedures to achieve high spatial resolution, and time-consuming sample preparation, e.g., placing individual NWs on cantilevers and electrically contacting them using electron beam lithography, metal deposition, and lift-off.

Here, we demonstrate a new and simple experimental approach based on photoluminescence (PL) spectroscopy to investigate the effect of diameter variation on light absorption in GaAs NW arrays and single NWs. In PL experiments, light absorption can be limited to single NWs by focusing the incident laser light, provided the NW spacing is larger than the laser spot. Also, the detection of the PL emitted by single NWs is easily possible for a sufficiently high quantum efficiency. Very generally speaking, the NW PL spectra depend in both intensity and wavelength on the excitation density. Indeed, PL excitation spectroscopy, in which the wavelength of the exciting laser is varied, has been employed to study for (In,Ga)As-GaAs core-shell NW ensembles how photons of different energy are absorbed in different parts of the NWs for different scattering geometries[24]. Our key idea to assess the amount of light absorbed in GaAs NWs is that the nature of carrier recombination depends on the carrier density, which in turn is affected by the absorption of the exciting laser light. In other words, if for a different NW diameter the light of the laser used for excitation is absorbed more efficiently, this effect will be reflected in the PL spectra.

For this study, the use of GaAs NWs with high phase-purity is crucial. Typically, the PL spectra of GaAs NWs contain numerous transitions caused by polytypism, and their precise energy depends sensitively on the details of the crystallographic stacking, i. e. the lengths of segments of either zincblende or wurtzite polytype[25,26]. For such NWs, correlating the spectra with the photogenerated carrier density would be challenging. Previously, our group demonstrated the growth of GaAs NWs whose PL spectra exhibit only transitions related to zincblende GaAs[27]. Such GaAs NWs are the basis for the current investigation.

## 2. Results and Discussion

We investigate GaAs NWs that are passivated by an $Al_{0.3}Ga_{0.7}As$ shell followed by a GaAs outer shell to suppress oxidation as illustrated by the schematic shown as inset to **Figure 1**. Individual core/multi-shell NWs from samples with different NW diameters are shown in bird's-eye view secondary electron (SE) micrographs. The NW diameter was tuned through thermal annealing by adjusting the annealing time[29]. As shown in **Figure 1**, the diameter decreases approximately linearly with annealing time, based on measurements at the NW center for 15 NWs per sample. The NWs exhibit tapering both before and after annealing, which is a characteristic feature of the vapor-liquid-solid (VLS) growth[28] method used for their fabrication (see Methods). We note that for all annealing times the nominal diameter of the NW core is greater than 30 nm, for which quantum confinement can be neglected[29]. For light coupling, the total diameter is most relevant since the complex refractive indices $\tilde{n}=n+ik$ of GaAs and $Al_{0.3}Ga_{0.7}As$ do not differ too much at the wavelengths used for excitation ($\tilde{n}=3.85+i0.19$ and $\tilde{n}=3.6+i0.13$ for an excitation wavelength $\lambda$ of 633 nm, respectively) [31,32]. Additionally, the SE micrographs reveal a broadened top segment on the NWs, formed as the Ga droplet that drives VLS growth solidifies into GaAs, stopping further axial growth and leaving a thicker top region.

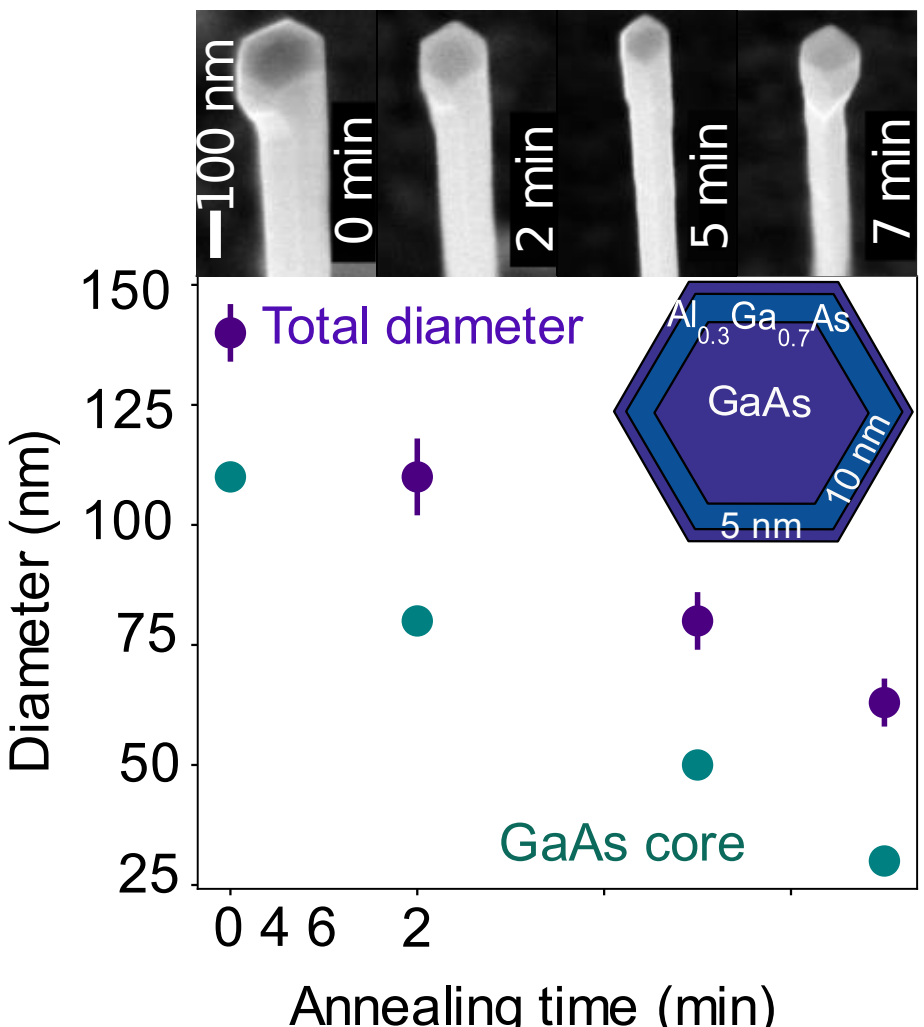

Figure 1. Thinning of the diameter of $GaAs/Al_{0.3}Ga_{0.7}As/GaAs$ core/multi-shell NWs by annealing of the GaAs core. (a) SE micrographs (viewing angle 15°) of individual NWs from each sample. The text labels indicate the annealing time. (b) Total NW diameter (average of 15 NWs per sample) over annealing time (purple data points). The cyan points correspond to the nominal diameter of the GaAs core. The inset shows a sketch of the NW multi-shell structure.

Power-dependent PL spectra were recorded from arrays of free-standing NW excited at 633 nm, with the corresponding micrographs shown in **Figure 2**. The arrays have a pitch of 0.7 µm, such that approximately 16 NWs are excited in each PL measurement (see Methods). The sample with the thinnest NWs annealed for 7 min was excluded from this analysis because its low fraction of vertical NWs (<20%) would result in the excitation of essentially single NWs.

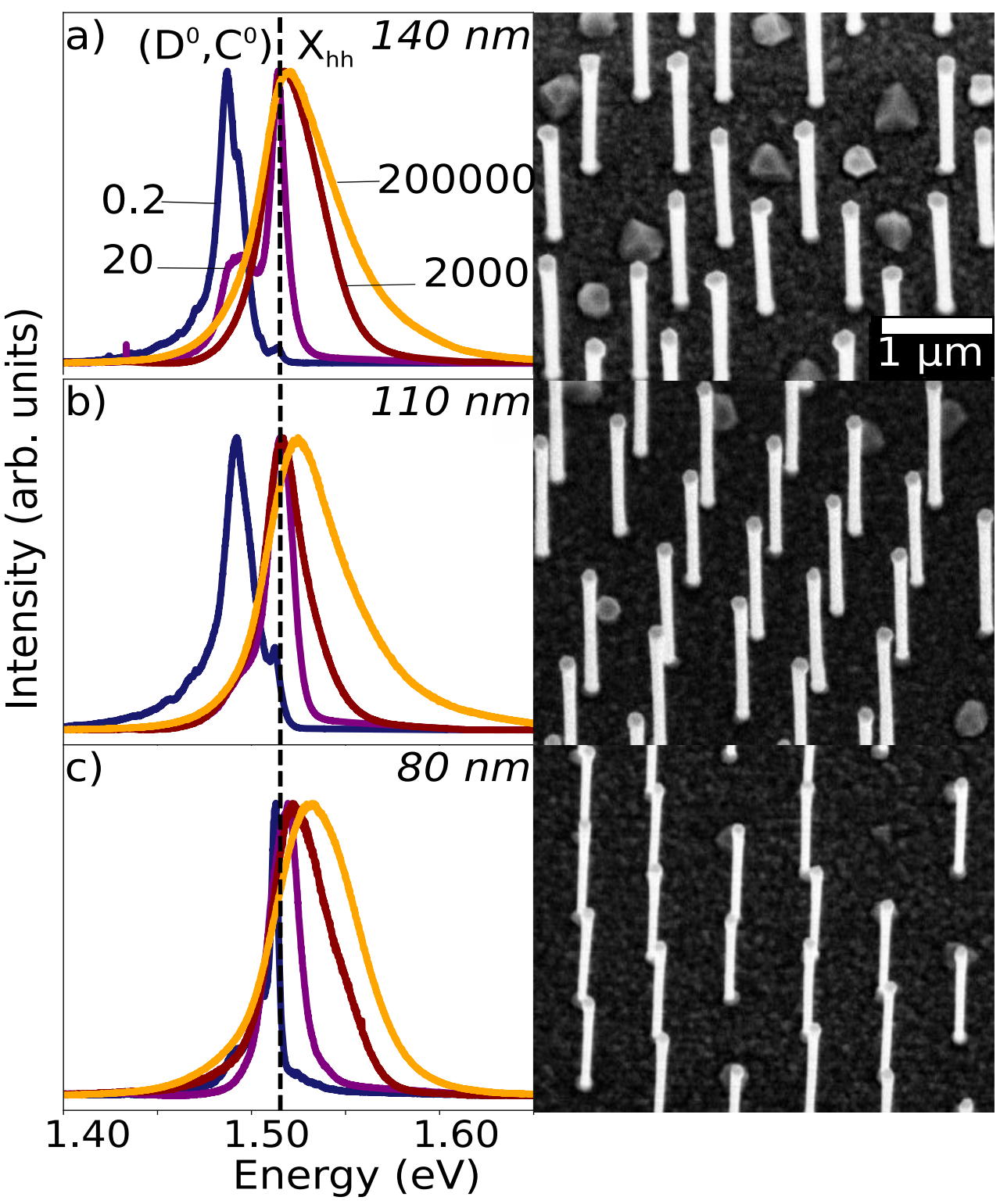

Figure 2. Normalized low-temperature (10 K) power-dependent PL spectra of GaAs/$Al_{0.3}Ga_{0.7}As$/GaAs NW arrays with different diameters with λ=633 nm and the corresponding bird's eye view SE micrographs. The excitation densities shown in (a) also apply to the spectra in (b) and (c) with units W/cm$^2$. The dashed line shows the $X_{hh}$ position of bulk GaAs.

The spectra of the as-grown NWs are presented in **Figure 2(a)** and display two emission lines originating from the donor-acceptor pair transition ($D^0$, $C^0$) with the shallow acceptor due to the incorporation of C, and the free-exciton transition $X_{hh}$. These emission lines are well known from high-quality epitaxial GaAs layers, and their identification in our NWs is discussed in detail in our previous publication focusing on NWs of a single diameter[27]. At low excitation densities, the ($D^0$, $C^0$) line dominates the spectra. As the excitation increases, the free exciton line becomes more intense until it dominates the spectra and the ($D^0$, $C^0$) transition saturates. This saturation takes place at progressively lower excitation densities with decreasing NW diameters [**Figures 2 (b)** and **(c)]**. Assuming a spatially uniform incorporation of C, the ($D^0$, $C^0$) transition should saturate at a certain, well-defined carrier density. The carrier density depends on photogeneration and thus the amount of light absorption but may be reduced by non-radiative recombination. If we assume that the non-radiative recombination rate does not depend on NW diameter, the earlier onset of saturation observed for thinner NWs indicates enhanced light absorption in these NWs. We will later discuss the validity of this assumption.

The dependence of light absorption on NW diameter can also be deduced by examining the evolution of the $X_{hh}$ transition with excitation density, as explained in the following. **Figure 3** shows the peak position for NW arrays and single NWs (measured in arrays with a pitch of 5 μm, i.e., larger than the laser spot size) with different diameter. Red data points correspond to the spectra from **Figure 2** acquired with an excitation wavelength of 633 nm, whereas blue data points belong to additional PL experiments with an excitation wavelength of 473 nm. The relevance of the second excitation wavelength will be discussed below. At the lowest excitation density, a red-shift with respect to the spectral position of the $X_{hh}$ line in bulk GaAs is observed for all NW diameters. This shift arises from the tensile strain exerted by the $Al_{0.3}Ga_{0.7}As$ shell on the GaAs core, which reduces the band gap[26,27].

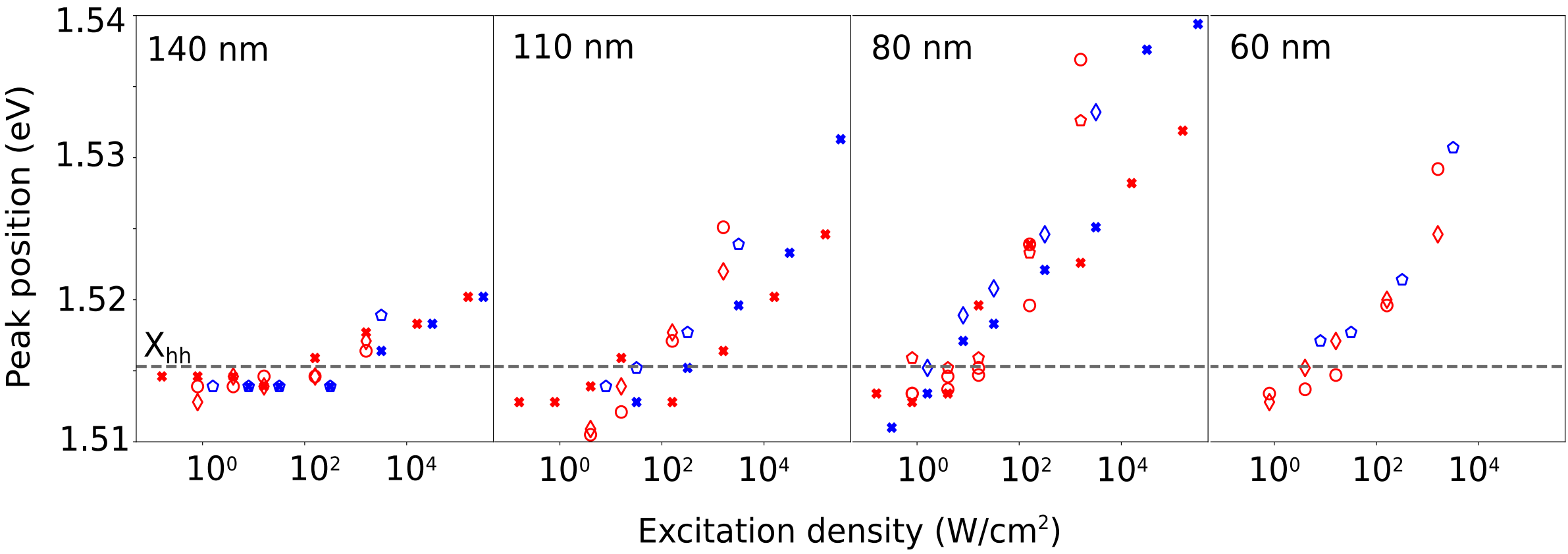

Figure 3. Dependence of the peak position on excitation density for the emission band at higher energy measured for $GaAs/Al_{0.3}Ga_{0.7}As/GaAs$ NW arrays and single NWs with different average NW diameters as displayed by the text labels. This emission band corresponds to the $X_{hh}$ transition at low excitation and to electron-hole-plasma recombination accompanied by a progressive band filling at high excitation. The red and blue data points refer to excitation wavelengths of 633 and 473 nm, respectively. Full symbols correspond to NW array data and open symbols to single NWs. Different shapes represent different single NWs. The dashed line indicates the position of the $X_{hh}$ line in bulk GaAs.

With increasing excitation density, the peak position shifts to energies that eventually exceed the band-gap of bulk GaAs. At the same time, the emission band exhibits a pronounced broadening of the high-energy side (see **Figure 2**). The blueshift and broadening signify a transition from excitonic to electron-hole-plasma recombination accompanied by a progressive band filling[27]. Similar to the saturation of the ($D^0$, $C^0$) line, this transition occurs at a well-defined carrier density (the Mott density) where all excitons are ionized, characterized by a screening length approximately equal to the free exciton radius. Clearly, the blueshift seen in **Figure 3** is more pronounced and sets in at lower excitation densities for smaller NW diameters, consistent with stronger absorption in thinner NWs.

The comparison of NW arrays and single NWs with the same diameter is partly handicapped by the fact that for single NWs, it proved to be difficult to collect reliable data for very low and very high excitation densities. In the former case, an optimum laser focus cannot easily be established because of the long integration times, and in the latter case, heating becomes a significant issue, manifested by a reduction in PL intensity and a sudden redshift of the emission. These difficulties also apply to the array with a NW diameter of 60 nm, where the low yield results essentially in the excitation of single NWs. Nevertheless, the data presented in **Figure 3** for the NW diameters of 80–140 nm reveal a larger blueshift for single NWs as compared to the NW arrays for the same excitation densities, suggesting correspondingly stronger absorption. This result can be understood from published simulations of the light interaction with GaAs NW arrays.[8] These simulations show that the NW absorption cross section increases with pitch up to a value corresponding to the NW length. For the present NW length of about 4 μm, the single NWs within the array with a pitch of 5 μm are thus expected to absorb more light than those in the arrays with a pitch of 0.7 μm, in agreement with our experimental results.

To directly display the diameter dependence of the light absorption in our single GaAs NWs, **Figures 4(a)** and **(b)** show the shift of the PL peak position versus the total NW diameter for the two excitation wavelengths. Data for single NWs with a GaAs core of about 110 and 50 nm diameter and an $Al_{0.1}Ga_{0.9}As$ shell of 30 nm thickness are shown in addition to the samples discussed so far (purple-gradient data points). The stronger blueshift with decreasing NW diameter is obvious from this representation of the data for both wavelengths used. The blueshift (and thus photogenerated carrier density) is maximized for a NW diameter of 80 nm and decreases again for thinner NWs. This trend implies a corresponding dependence of light absorption on NW diameter if—as mentioned above—the non-radiative recombination rate is the same in all samples. In principle, one could conceive that this rate actually varies with diameter and that this variation is the main origin of the trend observed in **Figures 4(a)** and **(b)**. Possible sources of non-radiative recombination are point defects in the GaAs core and at the interfaces to the passivating (Al, Ga)As shell. Interface recombination should be more relevant for thinner NWs but those exhibit actually a higher carrier density. Alternatively, the point defect density both in the bulk and at the interface could be affected by the annealing employed for NW thinning. However, that these changes give rise to the observed non-monotonic trend versus diameter is rather unlikely. Furthermore, the two samples with different core-shell configurations (purple data points) nicely fit into the overall trend, as expected for a photonic effect. Therefore, the most plausible interpretation is that the data of **Figures 4(a)** and **(b)** reveal a diameter dependence of light absorption in single NWs.

In order to verify this interpretation, we carried out simulations of light absorption enhancement in single GaAs NWs with a circular cross section and a length of 3.8 μm standing on a Si substrate and excited by a plane wave with λ=473 nm and 633 nm. The results are presented as solid lines in **Figures 4(c)** and **(d)**. For comparison with the experimental data, we use for all these plots the equivalent disk diameter D*, defined such that the area of the corresponding circular cross-section equals the hexagonal area of the experimental NWs. The simulations exhibit a very similar dependence on D*, but with the maximum at a NW diameter of about 100 nm for λ=633 nm and about 70 nm for λ=473 nm. Indeed, it is expected that light absorption peaks at a particular wavelength-diameter combination that induces a resonant mode. At the same time, the experimental data show no systematic difference between the two different excitation wavelengths within the scatter of the data and exhibit both a maximum at D*=84 nm. Possible explanations for this discrepancy and the small differences in the position of the maximum between simulations and experiments include a variation of the NW diameter within the arrays, which would essentially flatten and broaden the dependence of the absorption on the wavelength/diameter ratio, and details of the real NW geometry. In fact, such differences between experimental data and simulations are a common feature in reports on light absorption in NWs[13-16] and likewise attributed to deviations in details between simulated and experimental conditions.

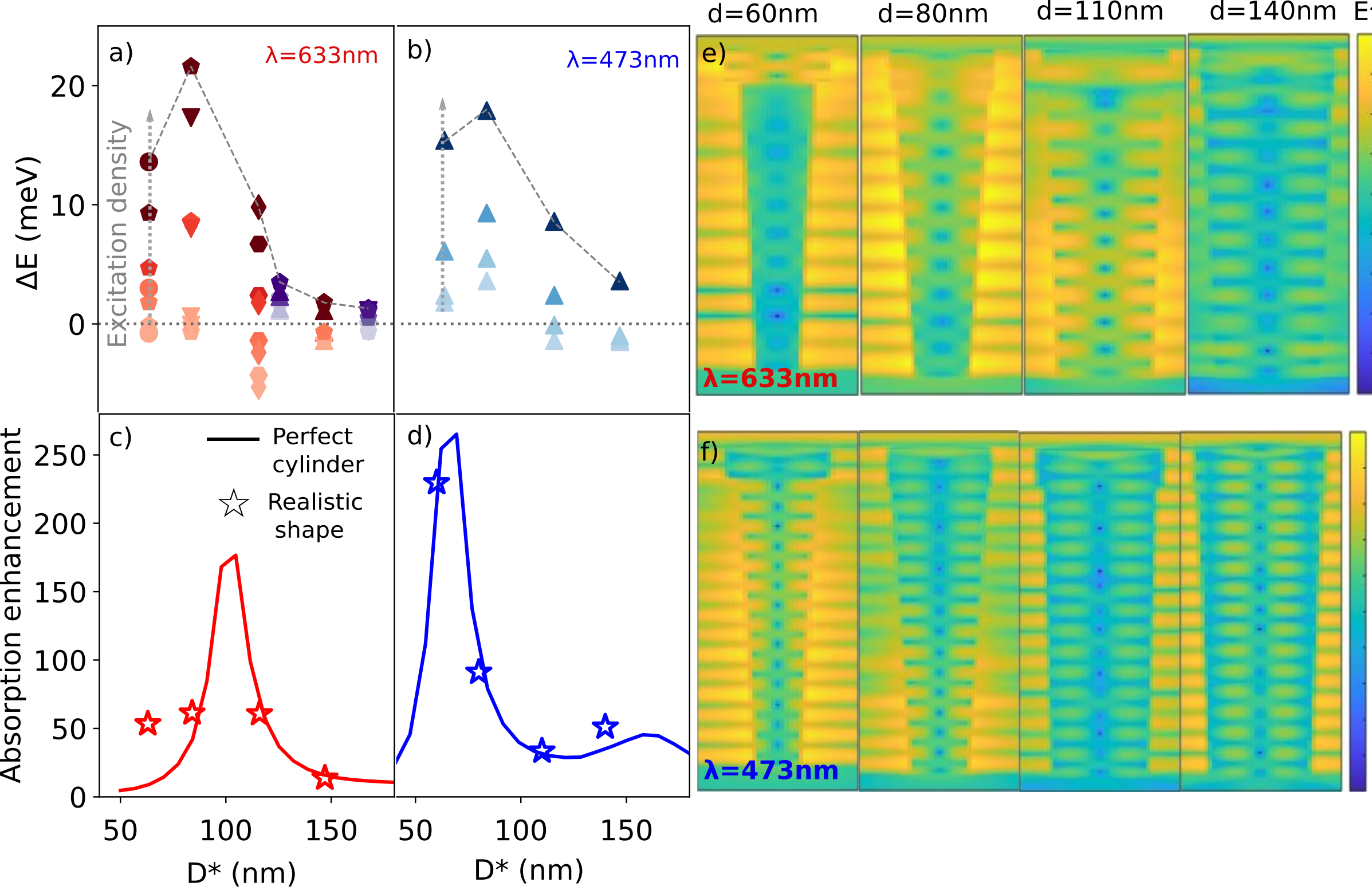

Figure 4 Evolution of the photoluminescence peak position with increasing excitation density as a function of equivalent disk diameter D* for excitation wavelengths of (a) 633 nm and (b) 473 nm. The same data acquired for single GaAs NWs is plotted in Fig. 3 in a different manner. Here, the increasing excitation density is displayed as a grading towards a darker shade in color and indicated by the dashed arrows. Different shapes represent different single NWs. The purple data points in (a) correspond to NWs with an $Al_{0.1}Ga_{0.9}As$ shell of 30 nm thickness that were not shown before. (c,d) Simulated absorption enhancement for GaAs NWs assuming either ideal cylindrical shapes (solid lines) or realistic core/shell geometries with hexagonal cross-section, inverse tapering, and a widened top segment (star symbols), for the same excitation wavelengths as in (a,b). (e,f) Log-scale maps of the squared electric field along the NW axis for realistic geometries and the two excitation wavelengths.

To explore possible reasons for the discrepancy found here, we performed refined simulations taking into account a more realistic NW geometry. In particular, the micrographs in **Figure 1(a)** show that the NWs exhibit at the top a widened segment resulting from the Ga droplet consumption after the core growth. The diameter and length of this top segment were measured for 15 NWs from each sample, and we used the average values summarized in **Table 1** for the refined simulations. Also, the table includes the diameters determined at the top and bottom of the NWs revealing the inverse tapering that was included in the simulations. In addition, the hexagonal cross-section of the NWs was taken into account. The results are plotted as star-shaped data points in **Figures 4(c)** and **(d)**. We restricted these simulations to the experimental measurements since the diameter of the top segment does not scale linearly with the main diameter and can thus not easily be extrapolated to a full diameter dependence of the absorption enhancement. Compared to simulations assuming perfectly cylindrical NWs, the realistic geometry increases for λ=633 nm the absorption enhancement for small diameters, whereas there is basically no change for diameters above 100 nm. For λ=473 nm, the two types of simulation agree very well.

**Table 1.** Dimensions of the widened NW top segments resulting from the Ga droplet consumption. 15 NWs per sample were measured. Diameters are expressed as equivalent disk diameters D* that results in a circle of the same area as the experimental hexagonal NW cross section. Indicated are the mean NW diameter D* measured at central height, bottom and top of the NWs, mean NW length $L_{NW}$, the mean diameter of the top segment $D^*_s$ and the mean segment length $L_s$ with their corresponding standard deviations.

.

| D* [nm] | $D^*_{Bottom}$ [nm] | $D^*_{Top}$ [nm] | $L_{NW}$[μm] | $D^*_s$ [nm] | $L_s$[nm] |
|---|---|---|---|---|---|
| 63 ± 5 | 53± 6 | 84± 3 | 3.94 ± 0.19 | 117 ± 6 | 385 ± 74 |
| 84 ± 6 | 62± 5 | 105± 4 | 2.94 ± 0.14 | 114 ± 6 | 464 ± 75 |
| 116 ± 8 | 106± 3 | 126± 5 | 3.74 ± 0.13 | 140 ± 7 | 435 ± 47 |
| 147 ± 6 | 137± 3 | 168± 4 | 3.62 ± 0.17 | 176 ± 17 | 550 ± 85 |

The influence of the realistic NW geometry on light absorption can be understood better on the basis of the data presented in **Figures 4(e)** and **(f)**. Maps of the square of the electric field, i. e. a parameter that is proportional to the light intensity, are depicted for all simulated realistic NW geometries. For λ=633 nm, the light intensity in the two thinnest NWs is clearly larger in the broadened top segment compared to the rest of these NWs but also compared to the thicker NWs. This observation holds in particular for the very thinnest NW and can be explained by the fact that the difference between the top and center diameters is larger for thin NWs. In contrast, under 473 nm excitation, the top segment does not exhibit significantly larger field strengths. Hence, the differences between the two types of simulations in **Figures 4(c)** and **(d)** are explained by the enhanced coupling of light with λ=633 nm to the widened top segment of thin NWs. Furthermore, we note that the maps in **Figures 4(e)** and **(f)** do not show any pronounced effect of the (Al,Ga)As shell whereas previously such shells were found to enhance absorption[30]. However, in our case the shell is only 10 nm thin, and the enhancement was reported only for larger shell thicknesses.

Both the experimental data in **Figures 4(a)** and **(b)** and the simulations in **Figures 4(c)** and **(d)** show that light absorption is stronger in NWs of diameter below about 100 nm compared to thicker NWs. In particular, the refined simulations for λ=633 nm suggest that the maximum in absorption enhancement is shifted to lower diameters than in the case of cylindrical NWs, in the direction where the experimental maximum is found. More generally, the comparison between simulations and the experimental data is limited by the fact that fairly sharp maxima are seen in the simulations of cylindrical NWs but experimental data and refined simulations are available only for selected diameters. For example, the simulations for λ=473 nm suggest a maximum around 70 nm, i. e. between the experimental data points. Furthermore, the refined simulations demonstrate that the precise combination of top segment diameter, NW diameter at the center, and wavelength has a strong influence on light absorption. Other details of the actual NW geometry such as the exact NW length may also affect light coupling. Therefore, we conclude that experimental results and simulations are in satisfactory agreement. This finding confirms that the trends in the experimental PL data of **Figures 2** and **3** indeed are unaffected by non-radiative recombination and reveal the dependence of light absorption on NW diameter.

## 3. Conclusion

In conclusion, we have demonstrated how PL spectroscopy can be employed to assess light absorption in single GaAs NWs and NW arrays with different diameters. Firstly, the ($D^0$, $C^0$) line saturates at a certain photogenerated carrier density depending only on C concentration. Secondly, the transition from excitonic to electron-hole pair recombination occurs at a higher and material-specific photogenerated carrier density. Our qualitative approach is much simpler than other experimental methods for the measurement of light absorption in single NWs, providing an easy route to experimentally analyze the engineering of leaky mode resonances by design of NW properties. Furthermore, it can be transferred to other material systems, provided their PL spectra exhibit features characteristic for a well-defined photogenerated carrier density. Therefore, our approach facilitates systematic experimental investigations of light absorption in single NWs beyond the diameter dependence of GaAs NWs studied here.

## 4. Experimental Section/Methods

GaAs NWs were grown by molecular beam epitaxy (MBE) on Si(111) substrates with a patterned $SiO_2$ mask using the Ga-assisted vapor-liquid-solid (VLS) mechanism according to our previously established growth protocol[27] with a V/III ratio of 4 and a substrate temperature of 640 °C. After core growth, the liquid Ga droplet was consumed by keeping only the As cell open. In a systematic series of samples, the NW diameter was reduced by congruent evaporation at 680 °C[29] for up to 7 min in the absence of any cell flux. Finally, the NWs were passivated by either a 10 nm thick $Al_{0.3}Ga_{0.7}As$ shell followed by a 5 nm thick GaAs cap layer to suppress oxidation, or by a 20 nm thick $Al_{0.1}Ga_{0.9}As$ shell. Both shells were grown at a V/III ratio of 10 and a substrate temperature of 500°C.

Power-dependent μ-PL spectra were obtained at nominally 10 K using either a He-Ne laser (wavelength λ=633 nm) or a diode-pumped solid-state (DPSS) laser (λ=473 nm) with spot diameters of ≈ 2 μm and ≈ 1 μm, respectively. The excitation density was controlled by neutral density filters and measured with a calibrated power meter.

Numerical simulations of light absorption in NWs were performed by directly solving Maxwell's equations in three dimensions by the FDTD method using the Lumerical™ software suite. The excitation source was a plane wave injected normally to the NWs with wavelengths λ=633 nm and 475 nm. The refractive index and extinction coefficient were taken from the literature for GaAs[31] and (Al,Ga)As[32]. Periodic boundary conditions were chosen in such a way that single NW absorption was studied. To determine the optical characteristics of the NWs, power monitors were placed above and below the vertical NWs to extract the reflectance R and transmittance T, and the absorptance (fraction of light absorbed inside the NW) was calculated using the relationship A=1-T-R[33]. The absorbed power ($P_{abs}$) is deduced as absorptance times the power ($P_{inc}$) of the source injected into the simulation region. The absorption enhancement is the ratio between $P_{abs}$ and $P_{inc}$, divided by the ratio between the physical NW cross section and the cell area[8]. Two sets of simulations were carried out. In the first one, cylindrical GaAs NWs were analyzed. In the second set, GaAs/$Al_{0.1}Ga_{0.9}As$/GaAs core-multi-shell NWs with hexagonal cross-section, inverse tapering, and a widened top segment were studied.

**Acknowledgements**
The authors thank Miriam Oliva for fruitful discussions, Manfred Ramsteiner for support with the PL setup, Claudia Herrmann for MBE maintenance, Anne-Kathrin Bluhm for SEM technical support, Abbes Tahraoui, Sander Rauwerdink and Walid Anders for substrate preparation as well as our colleagues Olaf Krüger, Mathias Matalla and Ina Ostermay from Ferdinand-Braun-Institut (Berlin) for e-beam lithography. We are grateful to Alexander Kuznetsov for a critical reading of the manuscript. We acknowledge funding from the Deutsche Forschungsgemeinschaft (DFG) under grant GE 2224/5-1.

## References

[1] L. Huang, L. Xu, D. A. Powell, W. J. Padilla, and A. E. Miroshnichenko, “Resonant leaky modes in all-dielectric metasystems: Fundamentals and applications,” *Phys. Rep.,* vol. 1008, pp. 1–66, Apr. 2023, doi: 10.1016/j.physrep.2023.01.001.

[2] J. Zhu *et al.*, “Optical Absorption Enhancement in Amorphous Silicon Nanowire and Nanocone Arrays,” *Nano Lett.*, vol. 9, no. 1, pp. 279–282, Jan. 2009, doi: 10.1021/nl802886y.

[3] J. Li, H. Yu, and Y. Li, “Solar energy harnessing in hexagonally arranged Si nanowire arrays and effects of array symmetry on optical characteristics,” *Nanotechnology*, vol. 23, no. 19, p. 194010, May 2012, doi: 10.1088/0957-4484/23/19/194010.

[4] D. Wu, X. Tang, K. Wang, Z. He, and X. Li, “An Efficient and Effective Design of InP Nanowires for Maximal Solar Energy Harvesting,” *Nanoscale Res. Lett.*, vol. 12, no. 1, p. 604, Dec. 2017, doi: 10.1186/s11671-017-2354-8.

[5] Z. Li, H. H. Tan, C. Jagadish, and L. Fu, “III–V Semiconductor Single Nanowire Solar Cells: A Review,” *Adv. Mater. Technol.,* vol. 3, no. 9, p. 1800005, Sep. 2018, doi: 10.1002/admt.201800005.

[6] N. I. Goktas, P. Wilson, A. Ghukasyan, D. Wagner, S. McNamee, and R. R. LaPierre, “Nanowires for energy: A review,” *Appl. Phys. Rev*., vol. 5, no. 4, p. 041305, Dec. 2018, doi: 10.1063/1.5054842.

[7] K. T. Fountaine, C. G. Kendall, and H. A. Atwater, “Near-unity broadband absorption designs for semiconducting nanowire arrays via localized radial mode excitation,” *Opt. Express*, vol. 22, no. S3, p. A930, May 2014, doi: 10.1364/OE.22.00A930.

[8] M. Heiss *et al.*, “III–V nanowire arrays: growth and light interaction,” *Nanotechnology*, vol. 25, no. 1, p. 014015, Jan. 2014, doi: 10.1088/0957-4484/25/1/014015.

[9] K. T. Fountaine, W. S. Whitney, and H. A. Atwater, “Resonant absorption in semiconductor nanowires and nanowire arrays: Relating leaky waveguide modes to Bloch photonic crystal modes,” *J. Appl. Phys*.**,** vol. 116, no. 15, p. 153106, Oct. 2014, doi: 10.1063/1.4898758.

[10] H. Bao and X. Ruan, “Optical absorption enhancement in disordered vertical silicon nanowire arrays for photovoltaic applications,” *Opt. Lett.*, vol. 35, no. 20, p. 3378, Oct. 2010, doi: 10.1364/OL.35.003378.

[11] L. Hu and G. Chen, “Analysis of Optical Absorption in Silicon Nanowire Arrays for Photovoltaic Applications,” *Nano Lett.*, vol. 7, no. 11, pp. 3249–3252, Nov. 2007, doi: 10.1021/nl071018b.

[12] Z. Gu, P. Prete, N. Lovergine, and B. Nabet, “On optical properties of GaAs and GaAs/AlGaAs core-shell periodic nanowire arrays,” *J. Appl. Phys.,* vol. 109, no. 6, p. 064314, Mar. 2011, doi: 10.1063/1.3555096.

[13] L. Cao, J. S. White, J.-S. Park, J. A. Schuller, B. M. Clemens, and M. L. Brongersma, “Engineering light absorption in semiconductor nanowire devices,” *Nat. Mater.***,** vol. 8, no. 8, pp. 643–647, Aug. 2009, doi: 10.1038/nmat2477.

[14] L. Cao *et al.*, “Semiconductor Nanowire Optical Antenna Solar Absorbers,” *Nano Lett.*, vol. 10, no. 2, pp. 439–445, Feb. 2010, doi: 10.1021/nl9036627.

[15] C. Colombo, P. Krogstrup, J. Nygård, M. L. Brongersma, and A. F. I. Morral, “Engineering light absorption in single-nanowire solar cells with metal nanoparticles,” *New J. Phys.*, vol. 13, no. 12, p. 123026, Dec. 2011, doi: 10.1088/1367-2630/13/12/123026.

[16] M. Y. Swinkels *et al.*, “Measuring the Optical Absorption of Single Nanowires,” *Phys. Rev. Applied*, vol. 14, no. 2, p. 024045, Aug. 2020, doi: 10.1103/PhysRevApplied.14.024045.

[17] M. D. Kelzenberg *et al.*, “Enhanced absorption and carrier collection in Si wire arrays for photovoltaic applications,” *Nat. Mater*, vol. 9, no. 3, pp. 239–244, Mar. 2010, doi: 10.1038/nmat2635.

[18] V. Protasenko, D. Bacinello, and M. Kuno, “Experimental Determination of the Absorption Cross-Section and Molar Extinction Coefficient of CdSe and CdTe Nanowires,” *J. Phys. Chem. B*, vol. 110, no. 50, pp. 25322–25331, Dec. 2006, doi: 10.1021/jp066034w.

[19] J. Giblin, F. Vietmeyer, M. P. McDonald, and M. Kuno, “Single Nanowire Extinction Spectroscopy,” *Nano Lett.*, vol. 11, no. 8, pp. 3307–3311, Aug. 2011, doi: 10.1021/nl201679d.

[20] M. P. McDonald, F. Vietmeyer, and M. Kuno, “Direct Measurement of Single CdSe Nanowire Extinction Polarization Anisotropies,” *J. Phys. Chem. Lett.*, vol. 3, no. 16, pp. 2215–2220, Aug. 2012, doi: 10.1021/jz3008112.

[21] R. Chatterjee, I. M. Pavlovetc, K. Aleshire, and M. Kuno, “Single Semiconductor Nanostructure Extinction Spectroscopy,” *J. Phys. Chem. C*, vol. 122, no. 29, pp. 16443–16463, Jul. 2018, doi: 10.1021/acs.jpcc.8b00790.

[22] P. Krogstrup *et al.*, “Single-nanowire solar cells beyond the Shockley–Queisser limit,” *Nat. Photonics*, vol. 7, no. 4, pp. 306–310, Apr. 2013, doi: 10.1038/nphoton.2013.32.

[23] S. A. Mann *et al.*, “Integrating Sphere Microscopy for Direct Absorption Measurements of Single Nanostructures,” *ACS Nano*, vol. 11, no. 2, pp. 1412–1418, Feb. 2017, doi: 10.1021/acsnano.6b06534.

[24] M. De Luca, “Addressing the electronic properties of III–V nanowires by photoluminescence excitation spectroscopy,” *J. Phys. D: Appl. Phys.*, vol. 50, no. 5, p. 054001, Feb. 2017, doi: 10.1088/1361-6463/50/5/054001.

[25] F. Bechstedt and A. Belabbes, “Structure, energetics, and electronic states of III–V compound polytypes,” *J. Phys.: Condens. Matter*, vol. 25, no. 27, p. 273201, Jul. 2013, doi: 10.1088/0953-8984/25/27/273201.

[26] A. Senichev *et al.*, “Electronic properties of wurtzite GaAs: A correlated structural, optical, and theoretical analysis of the same polytypic GaAs nanowire,” *Nano Res.*, vol. 11, no. 9, pp. 4708–4721, Sep. 2018, doi: 10.1007/s12274-018-2053-5.

[27] M. Oliva *et al.*, “Carrier Recombination in Highly Uniform and Phase-Pure GaAs/(Al,Ga)As Core/Shell Nanowire Arrays on Si(111): Implications for Light-Emitting

Devices," *ACS Appl. Nano Mater.*, vol. 6, no. 16, pp. 15278–15293, Aug. 2023, doi: 10.1021/acsanm.3c03242.

[28] H. Küpers *et al.*, "Diameter evolution of selective area grown Ga-assisted GaAs nanowires," *Nano Res.*, vol. 11, no. 5, pp. 2885–2893, May 2018, doi: 10.1007/s12274-018-1984-1.

[29] B. Loitsch *et al.*, "Tunable Quantum Confinement in Ultrathin, Optically Active Semiconductor Nanowires Via Reverse-Reaction Growth," *Adv. Mater.*, vol. 27, no. 13, pp. 2195–2202, Apr. 2015, doi: 10.1002/adma.201404900.

[30] A. Cretì, P. Prete, N. Lovergine, and M. Lomascolo, "Enhanced Optical Absorption of GaAs Near-Band-Edge Transitions in GaAs/AlGaAs Core–Shell Nanowires: Implications for Nanowire Solar Cells," *ACS Appl. Nano Mater.*, vol. 5, no. 12, pp. 18149–18158, Dec. 2022, doi: 10.1021/acsanm.2c04044.

[31] S. Adachi, "Optical dispersion relations for GaP, GaAs, GaSb, InP, InAs, InSb, Al *x* Ga1− *x* As, and In1− *x* Ga *x* As *y* P1− *y*," *J. Appl. Phys.,* vol. 66, no. 12, pp. 6030–6040, Dec. 1989, doi: 10.1063/1.343580.

[32] D. E. Aspnes, S. M. Kelso, R. A. Logan, and R. Bhat, "Optical properties of Al *x* Ga1− *x* As," *J. Appl. Phys.,* vol. 60, no. 2, pp. 754–767, Jul. 1986, doi: 10.1063/1.337426.

[33] F. Adibzadeh and S. Olyaee, "Optical absorption enhancement in vertical InP nanowire random structures for photovoltaic applications," *Opt Quantum Electron*, vol. 52, no. 1, p. 6, Jan. 2020, doi: 10.1007/s11082-019-2120-5.